\documentclass[review]{elsarticle}

\usepackage[utf8]{inputenc}
\usepackage[english]{babel}
\usepackage{amsmath,amssymb}
\usepackage{mathtools}
\usepackage{graphicx}
\journal{Journal of Molecular Spectroscopy}
\usepackage{fdsymbol}

\begin{document}

\begin{frontmatter}

\title{High-resolution spectroscopy of He${_2}^+$ using Rydberg-series extrapolation and Zeeman-decelerated supersonic beams of metastable He$_2$}

\author{Paul Jansen}
\author{Luca Semeria}
\author{Fr\'ed\'eric Merkt\corref{mycorrespondingauthor}}
\address{Laboratory of Physical Chemistry, ETH Zurich, CH-8093 Zurich, Switzerland}
\cortext[mycorrespondingauthor]{merkt@phys.chem.ethz.ch}

\begin{abstract}
Recently, high-resolution spectroscopy of slow beams of metastable helium molecules (He$_2^*$) generated by multistage Zeeman deceleration was used in combination with Rydberg-series extrapolation techniques to obtain the lowest rotational interval in the molecular helium ion at a precision of 18\,MHz [Jansen \emph{et al.} Phys. Rev. Lett. 115 (13) (2015) 133202], limited by the temporal width of the Fourier-transform-limited laser pulses used to record the spectra. 
We present here an extension of these measurements in which we have (1)  measured higher rotational intervals of He${_2}^+$, (2) replaced the pulsed UV laser by a cw UV laser and improved the resolution of the spectra by a factor of more than five, and (3) studied $M_J$ redistribution processes in regions of low magnetic fields of the Zeeman decelerator and shown how these processes can be exploited to assign transitions originating from specific spin-rotational levels ($N^{\prime\prime},J^{\prime\prime}$) of He$_2^*$.
\end{abstract}


\end{frontmatter}


\section{Introduction}
Translationally cold samples of molecules offer interesting perspectives for high-resolution spectroscopy. The long measurement times that are possible with such samples and the reduced Doppler widths are ideally suited for precision measurements of transition frequencies in molecules, and such measurements are beginning to be relevant in the context of tests of the standard model of particle physics and some of its extensions \cite{steimle2014}. Precision measurements in few-electron, light molecules such as H${_2}^+$, H$_2$ and He${_2}^+$ are used as tests of \emph{ab initio} quantum-chemical calculations which aim at an exact solution of the Schr{\"o}dinger equation and a rigorous determination of relativistic and quantum-electrodynamics (QED) corrections  \cite{korobov2006,korobov2008,piszczatowski2009,pachucki2010,korobov2014}. In these molecules, the velocity of the electrons is relatively low, and nonrelativistic quantum-electrodynamics turns out to be particularly successful. In this approach, the energy is expressed as a series expansion in powers of the fine-structure constant $\alpha$, which is a measure of the classical electron speed
\begin{equation}
E\left ( \alpha \right ) = \mathcal{E}^{(0)}+\alpha^2\mathcal{E}^{(2)}+\alpha^3\mathcal{E}^{(3)}+\alpha^4\mathcal{E}^{(4)}+\mathcal{O}\left (\alpha^5 \right ).
\end{equation}
The terms in different powers of $\alpha$ are associated with different contributions to the overall energy. $\mathcal{E}^{(0)}$ contains the Born-Oppenheimer energy including adiabatic and nonadiabatic interactions, while $\alpha^2\mathcal{E}^{(2)}$ and $\alpha^3\mathcal{E}^{(3)}$ represent the relativistic and leading-order QED corrections. Higher powers of $\alpha$ are associated with higher-order QED corrections. Recent calculations for the molecular hydrogen ion include relativistic and QED corrections up to terms proportional to $\alpha^6$ and report an accuracy of 2\,kHz for the first vibrational intervals of H${_2}^+$ \cite{korobov2008} and HD$^+$ \cite{korobov2014}. The most accurate calculations of the energies in H$_2$, HD, and D$_2$ include full corrections up to terms proportional to $\alpha^3$ as well as the dominant one-loop contribution of the $\alpha^4$ term \cite{piszczatowski2009,pachucki2010}. The reported uncertainties are less than 30\,MHz and the calculated and experimental results agree within this uncertainty \cite{liu2009,sprecher2011}. The best calculations of the rovibrational levels of He${_2}^+$ \cite{tung2012-1} have an accuracy of about 120\,MHz, sufficient to reproduce the energy-level structure measured in earlier experiments \cite{yu1987,yu1989-1}, although they do not include relativistic and radiative corrections.

Few-electron diatomic molecules present experimental challenges for high-resolution studies of their spectra, and experimental data on their energy-level structures are scarce. Indeed, the symmetric isotopomers do not have a permanent electric dipole moment, which implies that these species do not have a pure rotational nor a rovibrational spectrum. Spectroscopic data on the molecular helium cation are limited to rotational and vibrational transitions in the asymmetric $^3$He$^4$He$^+$ isotopomer reported by Yu \emph{et al.} \cite{yu1987,yu1989-1} and microwave transitions between highly excited vibrational levels of the electronic ground state and the lowest vibrational levels of the first electronically excited states in He${_2}^+$ by Carrington \emph{et al.} \cite{carrington1995}. The only experimental data available on the low-lying rovibrational levels of $^4$He${_2}^+$ have been obtained by photoelectron spectroscopy \cite{raunhardt2008} and from the Rydberg spectrum of He$_2$ using Rydberg-series extrapolation techniques \cite{ginter1980,ginter1984,raunhardt2008,sprecher2014,jansen2015}. 

The work presented in this article is devoted to measurements of the energy level structure of He${_2}^+$ by high-resolution spectroscopy of Rydberg states of He$_2$ and extrapolation of the Rydberg series 
\cite{raunhardt2008,sprecher2014,jansen2015}. The strength of our approach to study He${_2}^+$ relies on the facts that (1) its energy levels structure is obtained by extrapolation of allowed electronic transitions of He$_2$ in the ultraviolet (UV) range of the electromagnetic spectrum, (2) the initial state of He$_2$ we use, the metastable $a\,^3\Sigma_u^+$ state (called He$_2^*$ hereafter), can easily be generated in supersonic beams, (3) He$_2^*$ has a magnetic moment of two Bohr magneton, which makes it possible to decelerate He$_2^*$ beams to low velocities in the laboratory reference frame using the technique of multistage Zeeman deceleration \cite{vanhaecke2007,motsch2014}, and (4) the electronic spectrum of He$_2^*$ is well known and information on low-lying Rydberg states and fine-structure intervals facilitates the interpretation of spectra of high Rydberg states. We believe that, in the long term, these advantages will enable us to reach a higher precision and accuracy than currently possible in H$_2$ \cite{liu2009,sprecher2011,sprecher2013}.

The spectrum of He$_2$ has been investigated exhaustively after its first detection in 1913, independently by Curtis \cite{curtis1913} and Goldstein \cite{goldstein1913}. Most information about the Rydberg states of He$_2$ has been obtained with classical emission grating spectroscopy in the extensive measurements of Ginter and coworkers \cite{ginter1965,ginter1965-1,ginter1965-2,ginter1966,ginter1968,ginter1970-1,brown1971,ginter1983,ginter1984}. In addition, low-lying Rydberg states have been investigated using infrared emission \cite{hepner1956} and absorption \cite{gloersen1965} spectroscopy, laser-induced fluorescence spectroscopy \cite{miller1979}, Fourier-transform emission spectroscopy \cite{rogers1988,herzberg1986,focsa1998,hosaki2004}, laser absorption spectroscopy \cite{solka1987,kawakita1985,lorents1989,hazell1995},  optical heterodyne concentration-modulation spectroscopy \cite{li2010}, and infrared emission spectroscopy from proton-irradiated cryogenic helium gas \cite{brooks1988,tokaryk1995}. Highly accurate measurements of the fine structure in the lowest rotational states of He$_2^*$ ($\nu''=0$) have been performed by Lichten \emph{et al.} using molecular-beam radio-frequency (r.f.) spectroscopy \cite{lichten1974,vierima1975,lichten1978}. Bjerre and coworkers \cite{lorents1989,kristensen1990,hazell1995} employed laser-r.f. double-resonance spectroscopy to extend the measurements of the fine-structure intervals to higher rotational and vibrational states. Focsa \emph{et al.} \cite{focsa1998} performed a global fit on infrared and r.f. data to obtain a consistent set of molecular constants for the six lowest excited electronic states of He$_2$. 

The structure of this article is as follows: After an overview of current knowledge on He$_2^*$, on the triplet Rydberg states of He$_2$, and on the ground state of He${_2}^+$ in section~\ref{sec_He2}, we summarize our experimental approach in section~\ref{sec_exp}. The experimental results are presented in section~\ref{sec_results} and a brief summary is provided in the conclusions section.

\section{Energy levels of He$_2$ and He${_2}^+$: general considerations} \label{sec_He2}
The van-der-Waals interaction between two helium atoms in their $^1$S$_0$ ground state is extremely weak and gives rise to a very shallow potential-energy well for the ground state of He$_2$, with a depth in the order of $10^{-3}$\,cm$^{-1}$ \cite{cencek2012} and a single bound rovibrational state with a mean internuclear distance of almost 5 nm~\cite{grisenti2000}. In contrast, He${_2}^+$ in its $X^{+}\,^2\Sigma_u^+$ ground state is covalently bound, with a well depth of almost 2.5\,eV \cite{tung2012-1}. The strongly bound nature of the He${_2}^+$ ground state implies the existence of singlet and triplet Rydberg series of He$_2$. 
Many Rydberg states are known for He$_2$ that all belong to series converging on the $X^{+}\,^2\Sigma_u^+$ electronic ground state of He${_2}^+$~\cite{ginter1970,huber1979,ginter1984}. With the exception of the single bound level of the electronic ground state, all bound states of He$_2$ are Rydberg states, so that He$_2$ can be regarded as a Rydberg molecule~\cite{herzberg1987}. 

\subsection{Metastable helium molecules He$_2^*$}

The lowest Rydberg states of He$_2$, the $a\,^3\Sigma_u^+$ state, is metastable, with a calculated radiative lifetime of 18\,s~\cite{chabalowski1989}, because radiative decay to the ground electronic state is spin forbidden.  Its long lifetime and the ease with which it can be produced in electric discharges make He$_2^*$ an ideal initial state to study the electronic spectrum and the photoionization of He$_2$, as demonstrated in the numerous studies cited in the introduction. The triplet nature of He$_2^*$ gives rise to a magnetic moment of two Bohr magneton and thus to an electron-Zeeman effect that can be exploited to slow down supersonic beams of He$_2$ by multistage Zeeman deceleration \cite{motsch2014,jansen2015} (see also Sections~\ref{sec_exp} and~\ref{sec_results}).

\begin{figure}[bt]
\centering
\includegraphics[width=0.6\columnwidth]{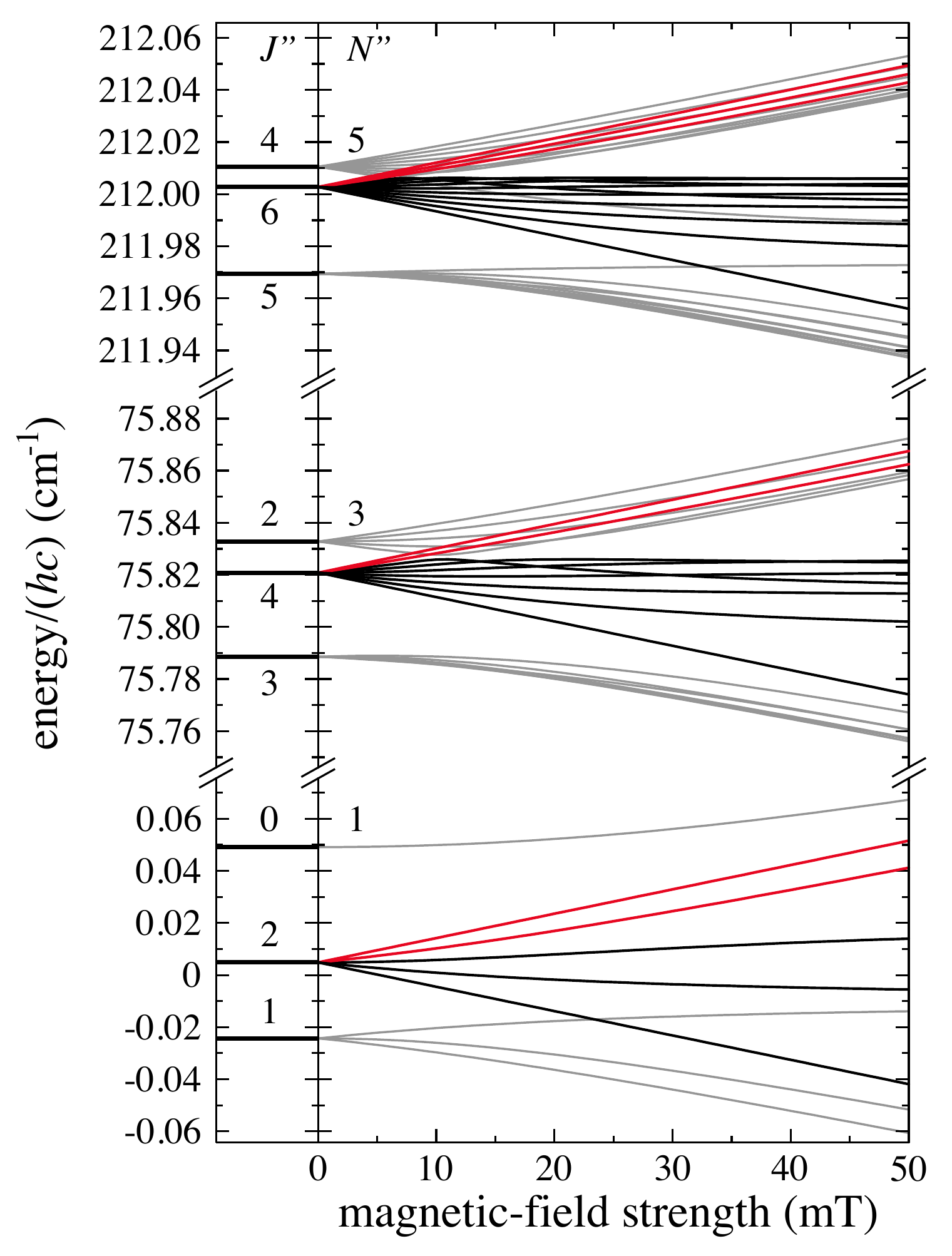}
\caption{Fine structure (left panel) and Zeeman effect (right panel) of the $N''=1,3$, and 5 states of He$_2^*$. The low-field-seeking magnetic sublevels of the $J''=N''+1$ manifold are shown in red and states of the $J''=N''$ and $N''-1$ are shown in grey. 
\label{fig:ZeemanShift}}
\end{figure}

The generalized Pauli principle requires the total wavefunction to be symmetric under exchange of the two bosonic $^4$He$^{2+}$ ($I=0$) nuclei, so that only rotational states for which the quantum number $N$ ($N$ is the quantum number associated with the total angular momentum excluding spin) is odd are allowed in states of $\Sigma_u^+$ symmetry, such as the $a\,^3\Sigma_u^+$ state of $^4$He$_2$ and the $X^+\,^2\Sigma_u^+$ state of He${_2}^+$. The rotational and fine structure in the vibrational ground state of He$_2^*$ can be described by an effective Hamiltonian~\cite{lefebvre-Brion2004,brown2003} appropriate to Hund's case (b) molecules in electronic states of $\Sigma$ symmetry

\begin{equation}
H = B_0 \vec{N}^2 - D_0 \vec{N}^4 + H_0 \vec{N}^6 + \tfrac{2}{3}\lambda_0 \left ( 3S_z^2-\vec{S}^2 \right ) + \gamma_0 \vec{S}\!\cdot\! \vec{N},
\label{eq:Hfs}
\end{equation}
where $B_0$ is the rotational constant, $D_0$ and $H_0$ are the quartic and sextic centrifugal distortion constants, $\lambda_0$ is the spin-spin interaction constant, $\gamma_0$ is the spin-rotation interaction constant, and $\vec{N}$ and $\vec{S}$ are the total angular momentum excluding spin and the total electron spin, respectively. The spin-spin and spin-rotation interactions split each rotational state $N$ into three fine-structure components with total angular momentum quantum number $J=N,N\pm 1$. Levels of the same $J$ value but $N$ values differing by 2 mix under the influence of the spin-spin interaction. Matrix elements of Eq.~\eqref{eq:Hfs} can be found in Ref.~\cite{brown2003}. The fine-structure splittings of the three lowest rotational states in He$_2$ ($a\,^3\Sigma_u^+$ $\nu=0$) are shown on the left-hand side of Fig.~\ref{fig:ZeemanShift}. 

To treat the effects of an external magnetic field, one needs two additional terms in the Hamiltonian
\begin{equation}
H_\text{Z} = -\frac{g_\text{e}\mu_\text{B}}{\hbar}\vec{S}\!\cdot\!\vec{B}-\frac{g_\text{R}\mu_\text{B}}{\hbar}\vec{N}\!\cdot\!\vec{B},
\label{eq:Hz}
\end{equation}
where $g_\text{e}\approx -2.00232$ and $g_\text{R}$ denote the electron and rotational $g$-factors, respectively. The second term on the right-hand side of Eq.~\eqref{eq:Hz} represents the coupling of the overall rotation of the molecule to the external magnetic field. This coupling is four orders of magnitude weaker than the coupling of the electron spin to the field and plays a negligible role for the deceleration experiments described below. On the right-hand side of Fig.~\ref{fig:ZeemanShift}, we show the eigenvalues of the combined Hamiltonian as a function of the magnetic-field strength for the three lowest rotational states in He$_2^*$. The Zeeman effect in higher rotational states is almost identical to that for $N=5$ because the fine structure at zero field only slowly changes with $N$ at high $N$ values and the rotational spacings are much larger than the Zeeman shifts, even for magnetic-field strengths of several Tesla \cite{motsch2014}.

\subsection{He${_2}^+$ and He$_2$ Rydberg states}

In the $X^+\,^2\Sigma_u^+$ electronic ground state, the rotational structure of He${_2}^+$ can also be described by Eq. (\ref{eq:Hfs}). However, only the spin-rotation interaction [last term on the right-hand side of Eq. (\ref{eq:Hfs})] contributes to the fine structure, because there is only one unpaired electron. The spin-rotation interaction in the $X^+\,^2\Sigma_u^+$ state of He${_2}^+$ splits each rotational state $N^+$ into two fine-structure states with $J^+=N^+\pm\tfrac{1}{2}$. In the following, we denote quantum numbers and molecular constants of the $a\,^3\Sigma_u^+$ state of He$_2$ and the $X^+\,^2\Sigma_u^+$ state of He${_2}^+$ with double primed symbols and a ``$+$'' superscript, respectively, to avoid confusion.
The splitting between the two $J^+$ states of a given $N^+$ value is given by $\gamma_0^+(N^+ +\tfrac{1}{2})$, where $\gamma_0^+$ is the spin-rotation coupling constant of the vibrational ground state of He${_2}^+$ ($X^+\,^2\Sigma_u^+$ $\nu^+=0$). 

Although $\gamma_0^+$ has not been measured for He${_2}^+$ yet, one can estimate it to be $\approx -3$\,MHz from the known spin-rotation coupling constant in $^3\text{He}^4\text{He}^+$ \cite{yu1989-1} and the fact that $\gamma_0^+$ scales as $\mu^{-1}$ \cite{brown1977}, where $\mu$ is the reduced mass of the molecule. To resolve the two fine-structure levels of the rotational states of He${_2}^+$ with $N^+$ values greater than 9 and 51, an experimental resolution of 25 and 150\,MHz, respectively, would be required.

\begin{figure}[bt]
\centering
\includegraphics[width=0.7\columnwidth]{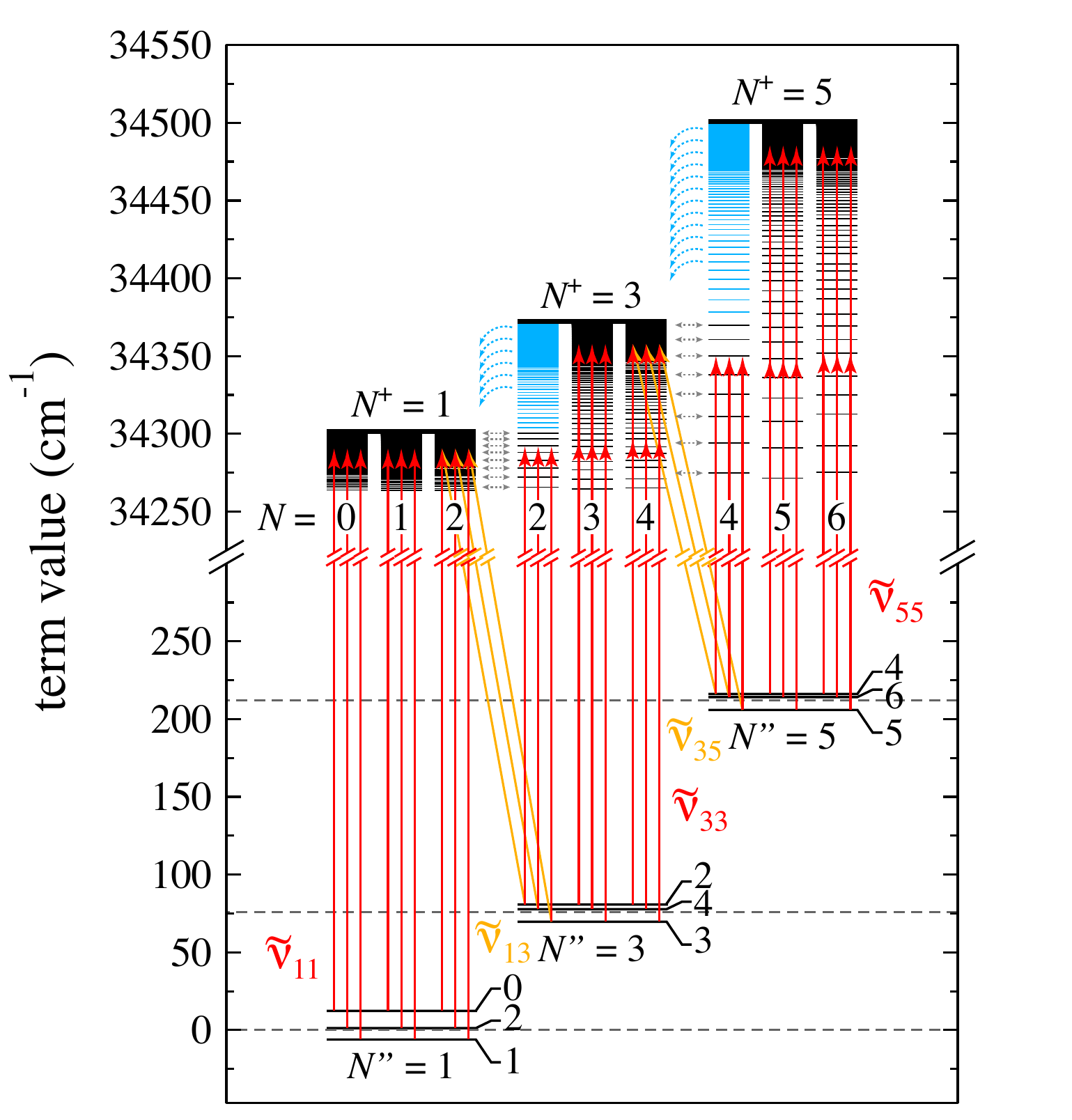}
\caption{Energy-level diagram showing the rotational levels of He$_2^*$ and the triplet $n$p Rydberg states of He$_2$ that converge to the three lowest rotational levels of He${_2}^+$. The positions of pure rotational levels of the metastable state are marked by dashed horizontal lines. The spin-rotation fine structure is exaggerated for clarity. Solid and dashed arrows indicate optically allowed transitions and channel interactions, respectively. Rapidly autoionizing levels are drawn in blue. 
\label{fig:energy_levels}}
\end{figure}

The energy-level structure of low-lying Rydberg states of He$_2$ is adequately described by Hund's angular-momentum coupling case (b) and thorough analyses of many of these low-lying states have been reported (see references cited in the introduction). Rydberg states of high principal quantum number are more conveniently described by Hund's angular-momentum coupling case (d). The Rydberg series of He$_2$ that can be accessed from He$_2^*$ by single-photon excitation are $np\sigma$ $^3\Sigma_g^+$ and $np\pi$ $^3\Pi_g^{\pm}$ in Hund's case (b) notation and $npN^+_N$ in Hund's case (d) notation. The two angular-momentum coupling schemes are related by a unitary angular-momentum frame transformation with which rotational-electronic interactions are treated in the realm of multichannel quantum-defect theory \cite{jungen2011}. The high-$n$ states we use to extrapolate the series and determine the rotational energy-level structure of He${_2}^+$ are best described in Hund's angular-momentum coupling case (d), as illustrated schematically in Fig.~\ref{fig:energy_levels}. Three $np$ series converge on each rotational level $N^+$ of He${_2}^+$, with $N$ values of $N^+$ and $N^+\pm 1$ ($\vec{N}= \vec{N^+}+\vec{\ell}$, with $\ell =1$ for $p$ series). The series with $N=N^+$ have pure Hund's case (b) $np\pi$ $^3\Pi_g^{-}$ character and those with $N=N^+\pm 1$ and $N>1$ have mixed $np\sigma$ $^3\Sigma_g^+$ and $np\pi$ $^3\Pi_g^{+}$ character.

Because $N$, and not $N^+$, is the good quantum number, levels of the same $N$ value that converge on different rotational states of He${_2}^+$ interact, giving rise to spectral perturbations below and to rotational autoionization above the $N^+=N-1$ thresholds. 
These interactions, indicated by horizontal arrows in Fig.~\ref{fig:energy_levels}, couple series differing in $N^+$ by 2 and need to be accounted for in the extrapolation of the Rydberg series. The best way to do so is by using multichannel quantum-defect theory as implemented by Jungen~\cite{jungen2011}. In our determination of the lowest rotational interval of He${_2}^+$ we extrapolated the series using the quantum defect parameters of the $n$p triplet states of He$_2$ reported by Sprecher \emph{et al.}~\cite{sprecher2014}. If the positions of the rotational levels of the cation are not known with sufficient precision, as is the case for the series converging to the $N^+=11$ and 13 ionic levels discussed in Section~\ref{sec_results}, the series limits can be extrapolated in first approximation using Rydberg's formula
\begin{equation}
hc\tilde{\nu}_{n\ell}=E_{\text{I}}\left (\text{He}_2^* \right ) + E_\text{rv}\left (\text{He}{_2}^+ \right )-\frac{hc\mathcal{R}_{\text{He}_2}}{\left ( n-\delta_\ell\right )^2},
\end{equation}
where $\tilde{\nu}_{n\ell}$ represents the spectral position of the Rydberg states of principal and orbital angular-momentum quantum numbers $n$ and $\ell$, respectively, and quantum defect $\delta_\ell$. The quantities $E_{\text{I}}(\text{He}_2^*)$, $E_\text{rv}(\text{He}{_2}^+)$ and $\mathcal{R}_{\text{He}_2}$ represent the adiabatic ionization energy of He$_2^*$, the rovibrational energy of the He${_2}^+$ ion core, and the mass-corrected Rydberg constant for He$_2$, respectively. Rydberg's formula adequately describes the noninteracting $npN^+_{N=N^+}$ series and also gives good extrapolation results for the other two series if states of very high $n$ values are used in the extrapolation.

The spin-spin coupling of the triplet Rydberg states of He$_2$ scales with $n^{-3}$ and becomes negligible at high $n$ values. The spin-rotation interaction is primarily that of the ion core, so that the fine structure converges to the spin-rotation splitting of He${_2}^+$ at high values of $n$ (see Fig.~6 of Ref.~\cite{haase2015}). 

The rotational selection rules for single-photon excitation from He$_2^*$ to $n$p$\sigma\,^3\Sigma_g^+$ and $n$p$\pi\,^3\Pi_g^\pm$ Rydberg states are given by $N-N''=0,\pm 1$. Combined with the $\Delta\ell=\pm 1$ selection rule for transitions between Rydberg states, the overall selection rule for single-photon excitation from the  $a\,^3\Sigma_u^+$ state to Rydberg levels converging to the $X^+\,^2\Sigma_u^+$ state is given by $\Delta N=N^+-N''=0,\pm 2$. Transitions from He$_2^*$ to $n$p Rydberg levels can therefore be unambiguously labeled as $N''_{J''}\rightarrow n\text{p}N^+_N$. Transitions to Rydberg states with $\Delta N=0,-2$, and $+2$ are referred to as Q($N''$)-type, O($N''$)-type, and S($N''$)-type transitions, respectively. The Q-type transitions are far stronger than the O- and S-type transitions because the He${_2}^+$ ion core is left with an electron hole of mainly s character after excitation of He$_2$ to $n$p Rydberg states \cite{willitsch2005}. However, O-type transitions can gain intensity through rotational channel interactions between levels converging on different rotational states of the ion discussed above and indicated by the horizontal dashed arrows in Fig.~\ref{fig:energy_levels}. The observation of these O-type transitions in He$_2$ is essential for the determination of the relative positions of energy levels in both the metastable state and ion ground state. 

\section{Experimental setup and procedure} \label{sec_exp}
\begin{figure*}[bt]
\centering
\includegraphics[width=1\columnwidth]{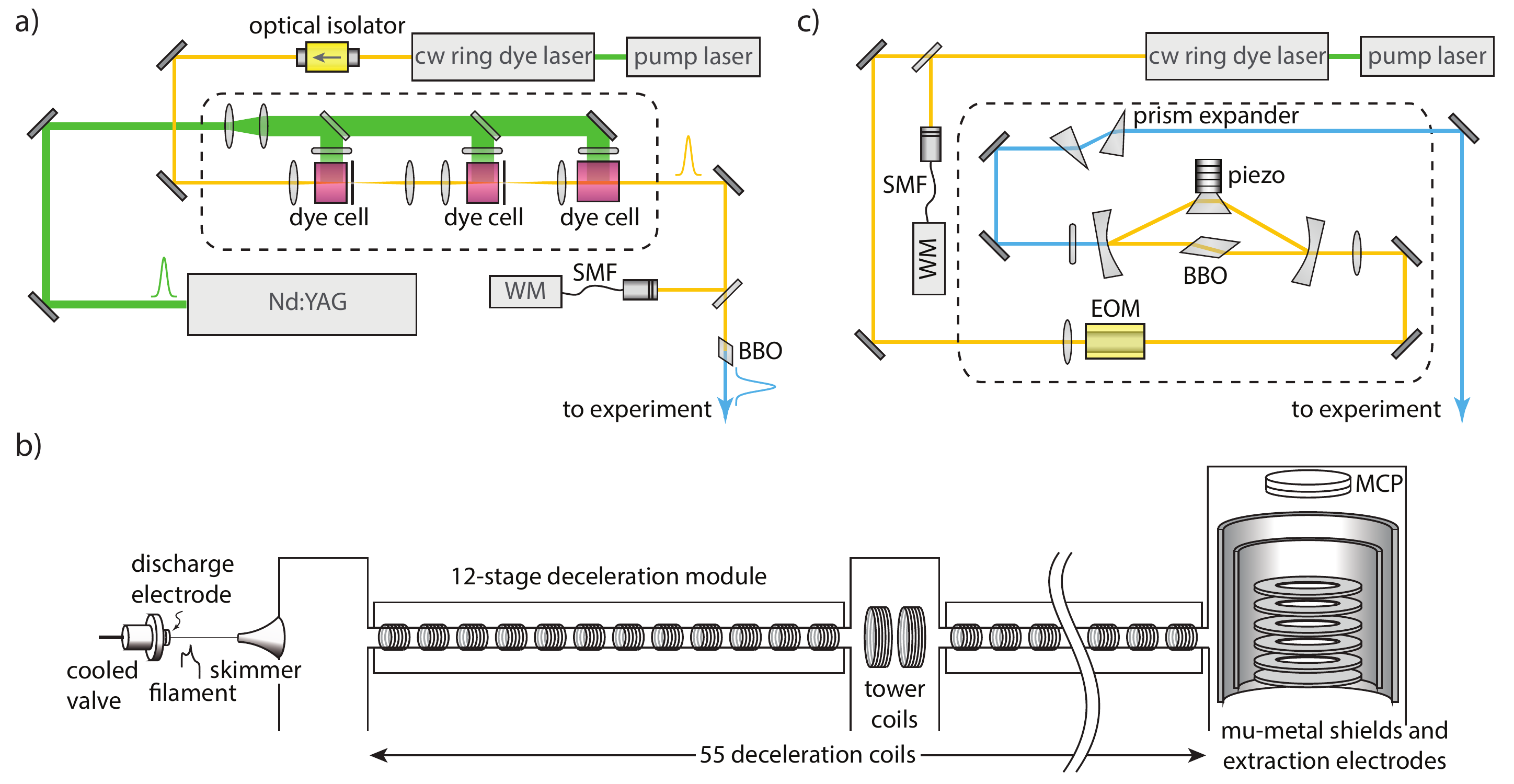}
\caption{Schematic representation (not to scale) of the experimental setup. a) pulsed-amplified ring-dye laser, b) vacuum setup showing the discharge source, the Zeeman decelerator, and the magnetically shielded photoexcitation region, and c) cw doubling of the output of a ring dye laser using a cavity. Nd:YAG, neodymium-doped yttrium aluminum garnet; WM, wave meter; SMF, single mode fiber; BBO, beta barium borate crystal; MCP, microchannel plate detector.
\label{fig:setup}}
\end{figure*}

A schematic view of the laser systems and the experimental setup is shown in Fig.~\ref{fig:setup}. In the experiments, we use both a pulsed UV laser system with a near Fourier-transform-limited bandwidth of 150\,MHz [depicted in Fig.~\ref{fig:setup}a)] and a continuous-wave (cw) single-mode UV laser with a bandwidth of 1.5\,MHz [depicted in Fig.~\ref{fig:setup}c)]. To generate the pulsed UV radiation, the cw output of a ring dye laser with a wavelength around 580\,nm is pulse amplified in dye cells pumped with the second harmonic of a neodymium-doped yttrium-aluminium-garnet (Nd:YAG) laser and frequency doubled in a beta-barium-borate (BBO) crystal. The pulse amplified fundamental radiation is frequency calibrated to an accuracy of 20\,MHz (1$\sigma$) with a wave meter and a fraction of the cw output of the ring dye laser is used to record the laser-induced-fluorescence spectrum of I$_2$, as desribed in Ref.~\cite{jansen2015}.  The difference between the frequencies of the cw and pulse-amplified outputs of the ring-dye laser is used to quantify the effects of the frequency chirp arising in the pulse-amplification process. The experiments involving cw UV radiation made use of the cw output of the ring laser, which was frequency doubled in an external cavity. In the experiments involving cw radiation, only the relative frequency was measured with an etalon at an accuracy of 5\,MHz. 

A supersonic beam of metastable helium molecules is produced in an electric discharge through an expansion of pure helium gas \cite{raunhardt2008} in a source chamber. The body of the valve can be cooled to temperatures of 77 and 10\,K, resulting in supersonic beams with velocities of approximately 1000 and 500\,m/s, respectively \cite{motsch2014}. The molecular beam is collimated with a skimmer before entering a second, differentially-pumped vacuum chamber that contains a 55-coil multistage Zeeman decelerator \cite{vanhaecke2007,motsch2014,wiederkehr2011}. 

The Zeeman decelerator exploits the Zeeman effect to manipulate the longitudinal velocity of the metastable helium molecules. When a He$_2^*$ molecule approaches an inhomogeneous magnetic field, it experiences a force that depends on its effective dipole moment. In a magnetic field, the $J''=2$ fine-structure component of the $N''=1$ rotational ground state of He$_2^*$ is split into five magnetic sublevels that are labeled with their value of $M_{J''}$, the quantum number associated with the projection of the total angular momentum vector $J''$ on the magnetic-field axis (see Fig.~\ref{fig:ZeemanShift}). The energy of two of the five magnetic sublevels increases as the magnetic-field strength increases. Molecules in these two states experience a force toward regions of low magnetic-field strength and are therefore referred to as ``low-field seekers''. Analogously, molecules in a state that displays a decrease in energy with increasing magnetic-field strength experience a force toward regions of high magnetic-field strength and are referred to as a ``high-field seekers''. When a low-field seeker approaches a magnetic field that is created by applying a current to a solenoid, part of its kinetic energy is converted into Zeeman energy and the molecule slows down. However, as soon as the low-field seeker crosses the region of maximum magnetic field, corresponding to the center of the solenoid, it gets accelerated again. In order to prevent this reacceleration, the magnetic field is switched off abruptly. By repeating this process many times and choosing the switch-off time of the current in the solenoids so as to maintain a phase-stable deceleration \cite{wiederkehr2010}, the molecules can be decelerated to any desired velocity. Because the magnetic field has to be switched off before the molecule leaves the coil, the maximum velocity that can be manipulated by a single coil is determined by the ratio of the length of the coil and the switch-off time of the magnetic field. For our experimental parameters (7.2\,mm, 8\,$\mu{s}$, 250\,A, and maximal on-axis field strength of 2\,T) this results in a maximum velocity around 700\,m/s. To obtain an initial velocity below this value, the valve body has to be cooled to 10\,K. 

The decelerator is segmented into three modules: two modules containing 12 coils and one module containing 31 coils. These modules are separated by pumping towers to guarantee a low background pressure in the decelerator. In addition, the modular design of the decelerator offers the flexibility to match the number of coils to the magnetic-moment-to-mass ratio of the species of interest. In order to maintain the magnetic quantization axis of the molecules as they traverse the region between the deceleration modules, the towers are equipped with solenoids as well, as explained by Wiederkehr \emph{et al.} \cite{wiederkehr2011}. 

\begin{figure}[bt]
\centering
\includegraphics[width=.4\columnwidth]{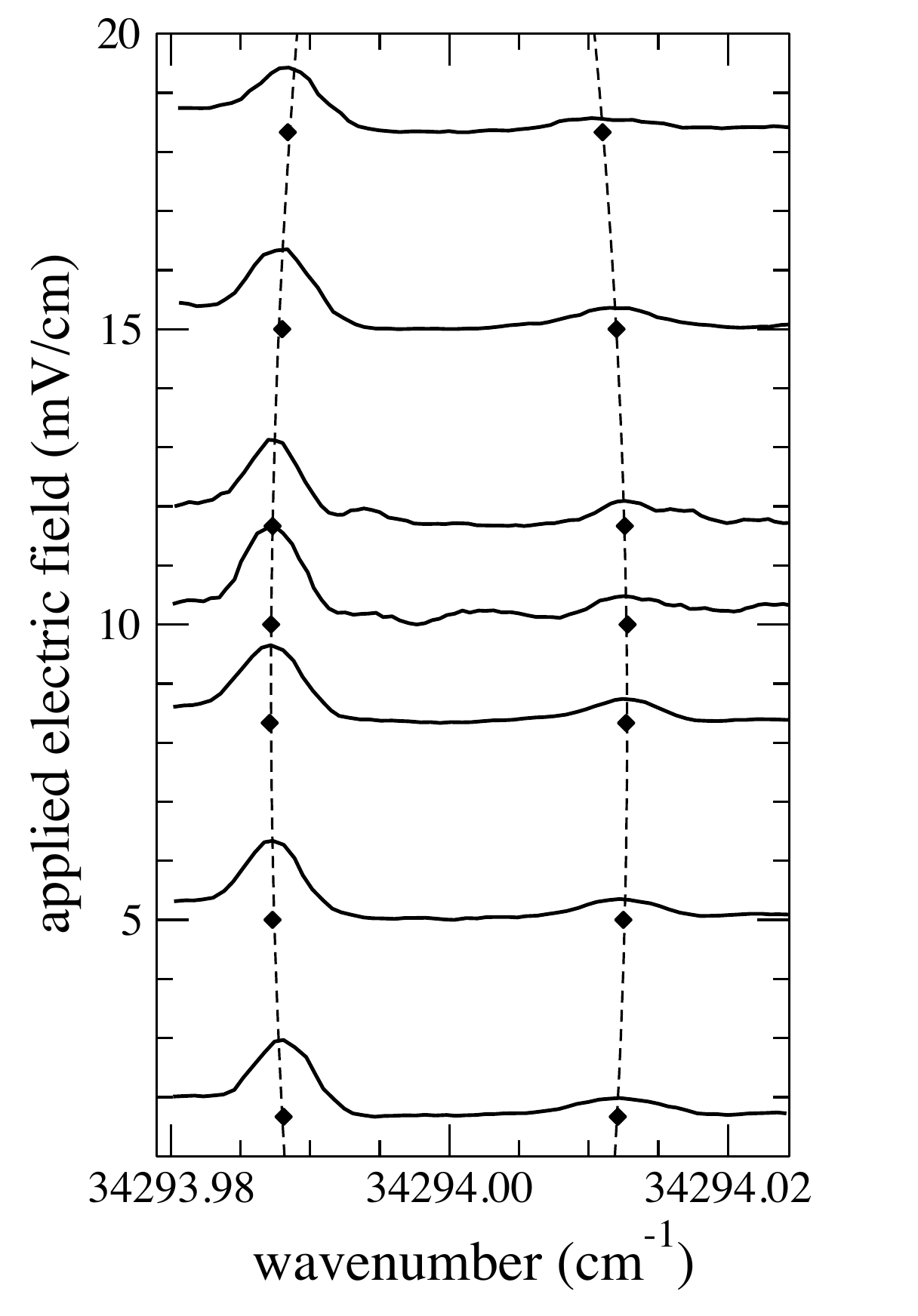}
\caption{Spectra of the $1_1\rightarrow 123\text{p}1_2$ transition (left-hand side) and $1_0\rightarrow 124\text{p}1_1$ transition (right-hand side)  of He$_2^*$ recorded in the presence of different dc electric fields. The vertical shift of the baseline of each spectrum corresponds to the applied field in units of mV/cm. The diamonds show the transition frequencies and the dashed lines represent fits using quadratic polynomials. In this way, the stray field could be reduced to below 1\,mV/cm.  
\label{fig:strayfieldcomp}}
\end{figure}

After about 1\,m of flight, the molecules enter a third vacuum chamber that is used for photoexcitation and detection. Approximately 60\,mm beyond the last coil of the decelerator, the molecular beam is intersected at right angles with the UV laser beam that is used to drive transitions to $n$p Rydberg states. The excitation region is surrounded by a cylindrically symmetric stack of electrodes for the application of ionization and extraction electric fields. A weak dc electric field is applied to the stack during photoexcitation in order to reduce stray electric fields to below 1\,mV/cm. The stray field is determined by recording spectra in the presence of different dc electric fields and fitting the observed Stark shifts to a quadratic polynomial, as illustrated in Fig. \ref{fig:strayfieldcomp}. In order to suppress stray magnetic fields, two concentric mu-metal tubes are used to shield the excitation region. For molecules decelerated to 120\,m/s, the current in the last coil is switched off 0.5\,ms before the molecules reach the excitation volume. The Rydberg states are ionized for detection by the application of a pulsed electric field which is also used to extract the ions toward a microchannel-plate (MCP) detector. A small electric field is applied to the stack shortly after photoexcitation but before the ionization pulse. This discrimination pulse separates prompt ions, produced by direct ionization or rapid autoionization, from ions produced by pulsed field ionization, based on their different arrival times on the MCP detector. The discrimination pulse also induces the field ionization of Rydberg states with $n\gtrsim200$, so that these states contribute to the prompt-ion signal.

\section{Results} \label{sec_results}
\begin{figure}[htb]
\centering
\includegraphics[width=0.7\columnwidth]{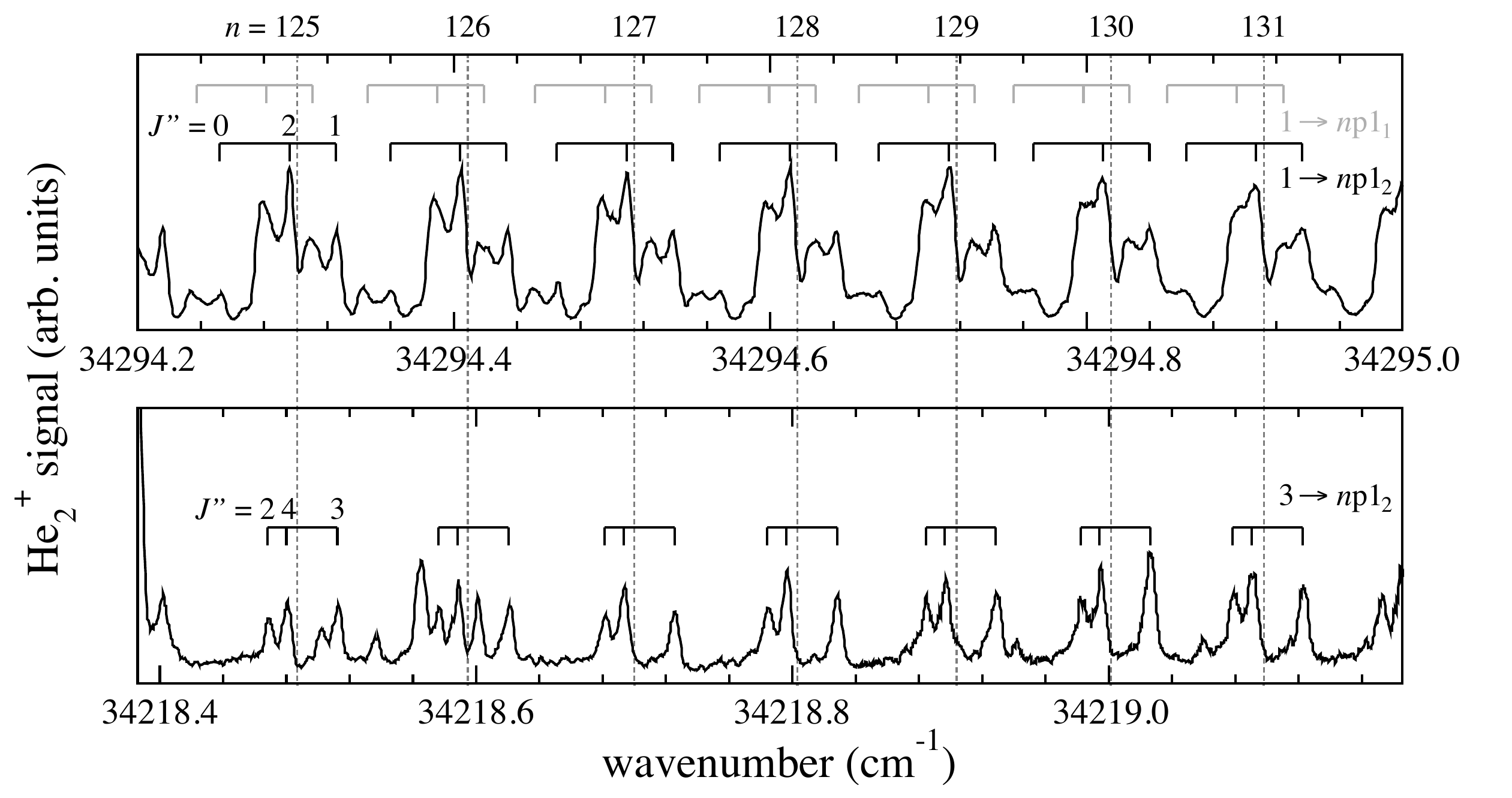}
\caption{Determination of the $\tilde{\nu}_{31}''$ interval between the $N''=1$ and 3 rotational levels of He$_2^*$ from combinational differences of transitions converging on the same $n\text{p}1_2$ Rydberg level. Vertical dashed lines indicate the centers of gravity of the transitions. 
\label{fig:combination_differences}}
\end{figure}

The procedure for obtaining the intervals between successive rotational states in He${_2}^+$ relies on the determination of the relative convergence limits of Rydberg series excited by Q-type and O-type transitions. 
This procedure has been used to extract the interval $\tilde{\nu}_{31}^+$ between the $N^+=1$ and $N^+=3$ rotational levels of He${_2}^+$ in Ref. \cite{jansen2015} and is briefly repeated here. $\tilde{\nu}_{31}^+$
corresponds to the difference between the convergence limits $\tilde{\nu}_{13}$ of the $3\rightarrow n\text{p}1_2$ series and $\tilde{\nu}_{33}$ of the $3\rightarrow n\text{p}3_3$ series (see Fig.~\ref{fig:energy_levels}). This interval can also be determined from the convergence limits $\tilde{\nu}_{11}$ of the $1\rightarrow n\text{p}1_{0-2}$ series and $\tilde{\nu}_{33}$ of the $3\rightarrow n\text{p}3_3$ series in combination with the interval $\tilde{\nu}_{31}''$ between the $N''=1$ and $N''=3$ rotational levels of He$_2^*$ and is given by $\tilde{\nu}_{13}^+=\tilde{\nu}_{31}''+\tilde{\nu}_{33}-\tilde{\nu}_{11}$. In order to derive the $\tilde{\nu}_{31}''$ interval, differences between transition wave numbers from $N''=1$ and $N''=3$ rotational levels to any member of the $n$p$1_2$ series can be taken. This procedure is illustrated in Fig.~\ref{fig:combination_differences} that shows the $1\rightarrow n\text{p}1_{1,2}$ and $3\rightarrow n\text{p}1_{2}$ Rydberg series around $n=128$  \cite{jansen2015}. To account for the triplet structure of the initial states, three Gaussian profiles have been fitted to the observed line shapes. The energy splitting between the line centers is fixed to the known fine-structure intervals \cite{lichten1974} and the relative intensities are assumed to correspond to the degeneracy factors $2J''+1$. The interval between the $N''=1$ and 3 rotational levels of He$_2^*$ was determined to be 75.8137(4)\,cm$^{-1}$ \cite{jansen2015}, in agreement with the value of 75.8129(3)\,cm$^{-1}$ derived from the rotational Hamiltonian and molecular constants reported by Focsa \emph{et al.}~\cite{focsa1998}. 

The convergence limits of the Rydberg series were derived by extrapolating transitions to Rydberg states of principal quantum numbers in the range $n=95-115$. This choice represents a compromise between states that are high enough in $n$ for the uncertainties in the quantum defects to have a negligible effect on the extrapolated energies and low enough not to be strongly affected by the Stark effect. The ionization thresholds $\tilde{\nu}_{11}, \tilde{\nu}_{13}$ and $\tilde{\nu}_{33}$ were determined to be 34301.20585(10), 34225.39234(10), and 34296.33037(7)\,cm$^{-1}$, respectively, with a systematic uncertainty of $1.4\times 10^{-3}$\,cm$^{-1}$ \cite{jansen2015}. The lowest rotational interval in He${_2}^+$ was thus determined to be 70.9380(6)\,cm$^{-1}$ with a total uncertainty of $6\times 10^{-4}$\,cm$^{-1}$ or 18\,MHz, which is less than the estimated uncertainty ($\approx 4\times 10^{-3}$\,cm$^{-1}$) of the most recent and precise theoretical value \cite{tung2012-1}.

\begin{figure}[bt]
\centering
\includegraphics[width=0.7\columnwidth]{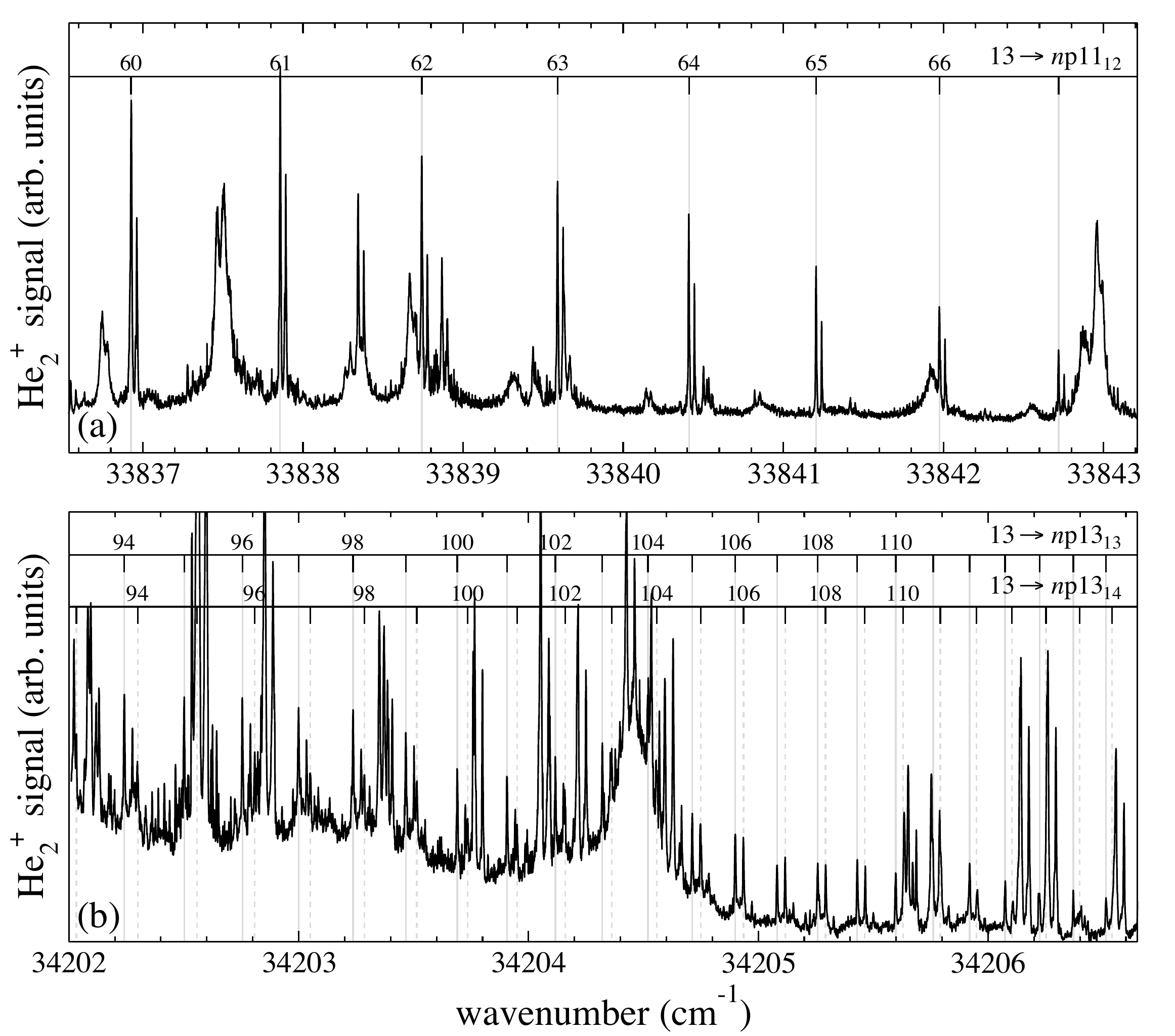}
\caption{Rydberg spectra of the (a) $13\rightarrow n\text{p}11_{12}$ series of He$_2$ in the range $n\approx 60-67$ and (b) $13\rightarrow n\text{p}13_{13,14}$ series in the range $n\approx 95-115$. The spectra also contain contributions from transitions belonging to $1\rightarrow n\text{p}1_{1,2}$ to $11\rightarrow n\text{p}11_{11,12}$ Rydberg series. 
\label{fig:Q13_O13}}
\end{figure}

Higher rotational intervals of He${_2}^+$ can be determined by recording Rydberg series that converge on higher-rotational states of the ion. As an example, spectra of the $13\rightarrow n\text{p}11_{12}$ Rydberg series in the range $n\approx 60-67$ and $13\rightarrow n\text{p}13_{13,14}$ Rydberg series in the range $n\approx 95-115$ are shown in Fig.~\ref{fig:Q13_O13}(a) and (b), respectively. The spectra also contain spectral features that correspond to transitions of the $1\rightarrow n\text{p}1_{1,2}$ to $11\rightarrow n\text{p}11_{11,12}$ series. The energy interval between the $N^+=11$ and 13 rotational states in He${_2}^+$ can be estimated 
by taking the difference between the series limits extrapolated using Rydberg's formula. Setting the uncertainty equal to the experimental linewidth, we determine this interval to be 346.988(6)\,cm$^{-1}$, as compared to the theoretical value of 346.977(4)\,cm$^{-1}$ \cite{tung2012-1}. Extrapolation of the $13\rightarrow n\text{p}11_{12}$ and $13\rightarrow n\text{p}13_{13,14}$ Rydberg series using MQDT should result in a more accurate determination of the $N^+=11$-to-13 rotational interval in He${_2}^+$. However, an accurate determination by MQDT requires the precise knowledge of higher rotational levels of He${_2}^+$, which is not available yet but can be obtained in a global analysis of all measured O-type and Q-type transitions to Rydberg states. Such an analysis is currently being performed, including series converging to rotational levels of He${_2}^+$ as high as $N^+=21$ \cite{semeria2016}.

\begin{figure}[bt]
\centering
\includegraphics[width=0.7\columnwidth]{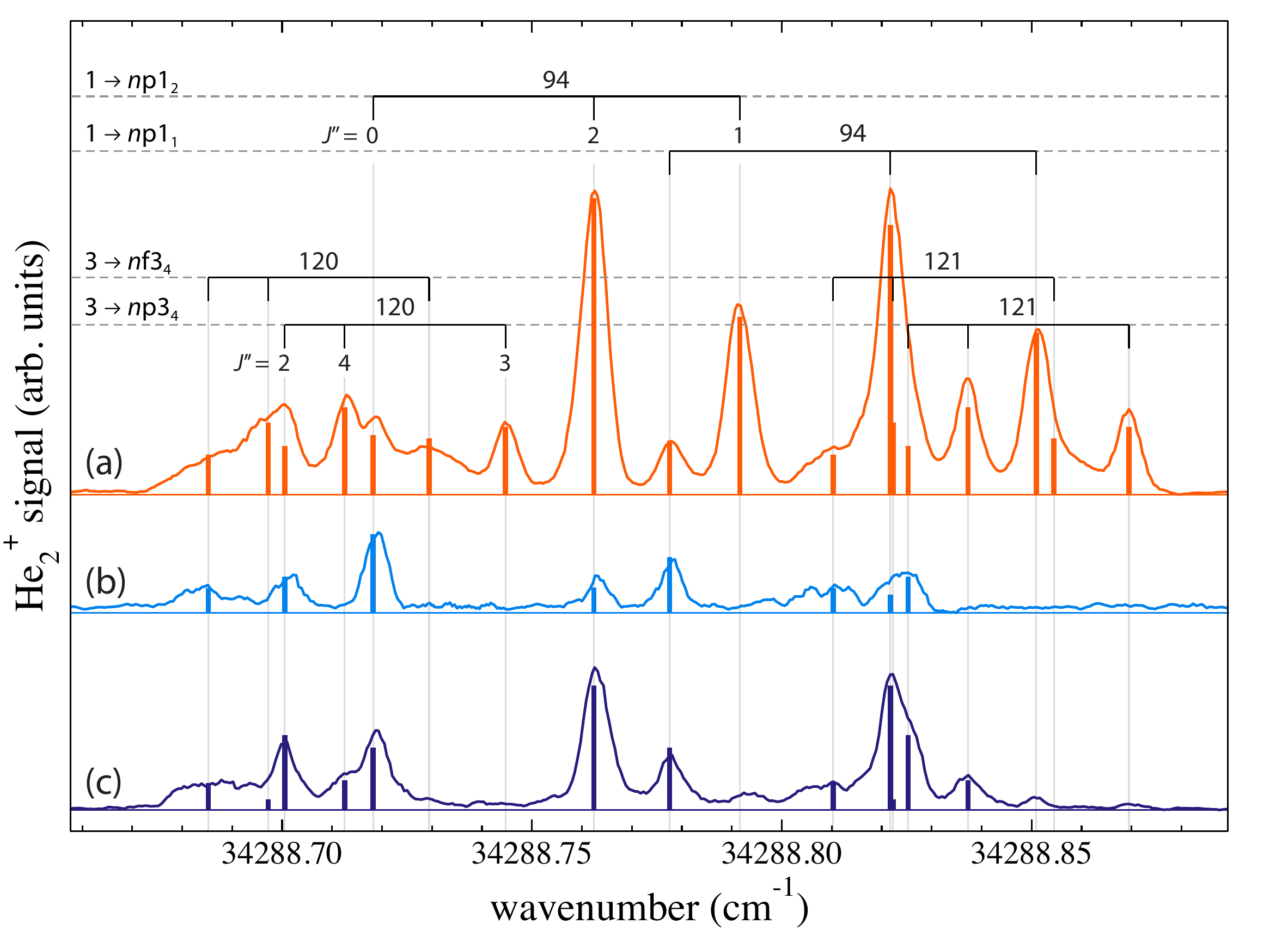}
\caption{Comparison between relative intensities of rotational fine structure components for spectra obtained from nondecelerated samples of He$_2^*$ [1000\,m/s, trace (a)], a spectrum obtained from a sample decelerated from 500 to 120\,m/s using three deceleration modules and two pumping towers [trace (b)], and a spectrum obtained from a sample decelerated from 500 to 335\,m/s using a single 31-stage deceleration module without any tower [trace (c)]. Vertical bars indicate expected relative line intensities based on the Zeeman energy of the different $M_{J''}$ states and assuming a complete redistribution among near-degenerate $M_{J^{\prime\prime}}$ levels. Note that for trace (c) the decelerated and undecelerated molecules are not completely separated spatially, giving rise to small features from the $J''=N''$ fine-structure components. 
\label{fig:dec_undec}}
\end{figure}

An important asset of the experiment is the use of a multistage Zeeman decelerator to reduce the velocity of the beam of metastable helium molecules and, consequently, systematic Doppler \textit{shifts} resulting from the velocity components parallel to the laser beam. In Fig.~\ref{fig:dec_undec}, traces (a) and (b) show part of the Rydberg spectrum of He$_2^*$ in the vicinity of the $1\rightarrow 94\text{p}1_{1,2}$ transition that was obtained using an undecelerated beam with a velocity of 1000\,m/s and a decelerated beam with a velocity of 120\,m/s, respectively. Many lines that are present in the spectrum recorded with the undecelerated beam are not observed in the spectrum obtained with the decelerated beam. This reduction in spectral congestion is a consequence of the spin-rotational state selectivity of the deceleration process. As expected from the Zeeman map of Fig.~\ref{fig:ZeemanShift}, the $J''=N''$ component is completely rejected from the beam, and transitions originating from this level are absent after deceleration. In addition, the $J''=2$ component carries approximately half the intensity of the $J''=0$ component for the Q(1)-type transitions and in the case of the Q(3)-type transitions, the $J''=4$ component is hardly visible in the spectrum and only the $J''=2$ component is observed. From the Zeeman map of He$_2^*$ shown in Fig.~\ref{fig:ZeemanShift}, one would expect that the relative intensities of the $J''=0$ and 2, and $J''=2$ and 4 fine-structure states would reflect the respective number of low-field seeking states, that is, one would expect ratios of 1:2 and 5:2, respectively, in contrast to the ratios of 2:1 and $>$20:1 observed experimentally. 

The apparent loss of molecules in the $J''=N''+1$ fine-structure component during the deceleration process can be understood in terms of a redistribution over all $M_{J''}$ states in regions of near-zero magnetic-field strength in the pumping towers. Although the pumping towers are equipped with coils that are pulsed as the molecules pass, the generated magnetic field is not strong enough in this case to maintain the magnetic quantization axis over the whole tower region, resulting in nonadiabatic losses. Assuming a complete redistribution over all $M_{J''}$ states every time the molecules traverse a tower, one expects the population of the $J''=2$ and 4 components to be suppressed by factors of $(5/2)^\tau$ and $(9/2)^\tau$, respectively, where $\tau$ is the number of towers the molecules traverse. The spectra in Fig.~\ref{fig:dec_undec} have been normalized with respect to the $J''=0$ fine-structure component of the transition to the 94$n$p$1_2$ Rydberg state and vertical bars indicate the expected intensities of the fine-structure components with respect to the $J''=N''-1$ state, assuming a complete redistribution over $M_{J''}$ states for the decelerated sample with $\tau=2$. Trace (c) in Fig.~\ref{fig:dec_undec} displays the same spectral region but was obtained using a molecular beam that was decelerated from 500 to 335\,m/s using a single 31-stage deceleration module without any tower ($\tau=0$). The relative intensities of the fine-structure components in this spectrum exactly matches the number of low-field seeking states. The observed changes of intensities resulting from $M_{J^{\prime\prime}}$ redistribution processes confirm the validity of the assumption of a statistical redistribution among the near-degenerate magnetic sublevels at each tower. These $M_{J^{\prime\prime}}$ redistribution processes actually turn out to be a powerful tool to assign transitions to specific initial $N^{\prime\prime}, J^{\prime\prime}$ levels.

Figure~\ref{fig:dec_undec} also indicates that no residual Doppler \textit{shifts} persist within the statistical uncertainty of the measurements. In principle, a reduction of the beam velocity results in a reduction of the Doppler \textit{width} of spectral lines and should therefore allow for a more precise determination of observed spectral positions. However, the pulse-amplification process results in a spectral resolution that is limited by the pulse width of the pumping laser. Assuming a Fourier-transform-limited Gaussian pulse with a pulse duration of 4\,ns, the frequency-doubled output of the laser has a bandwidth of about 150\,MHz. Because this bandwidth is larger than the Doppler width, no reduction of the line widths was observed for decelerated molecules. In order to observe an effect of the beam velocity on the Doppler width of the measured transitions, the bandwidth of the laser system has to be reduced below the Doppler linewidth. Because of the inherent limitations of a pulsed-laser system, this can be best achieved by replacing the pulsed laser by a cw laser, as shown in Fig.~\ref{fig:setup}c). 

\begin{figure}[bt]
\centering
\includegraphics[width=0.6\columnwidth]{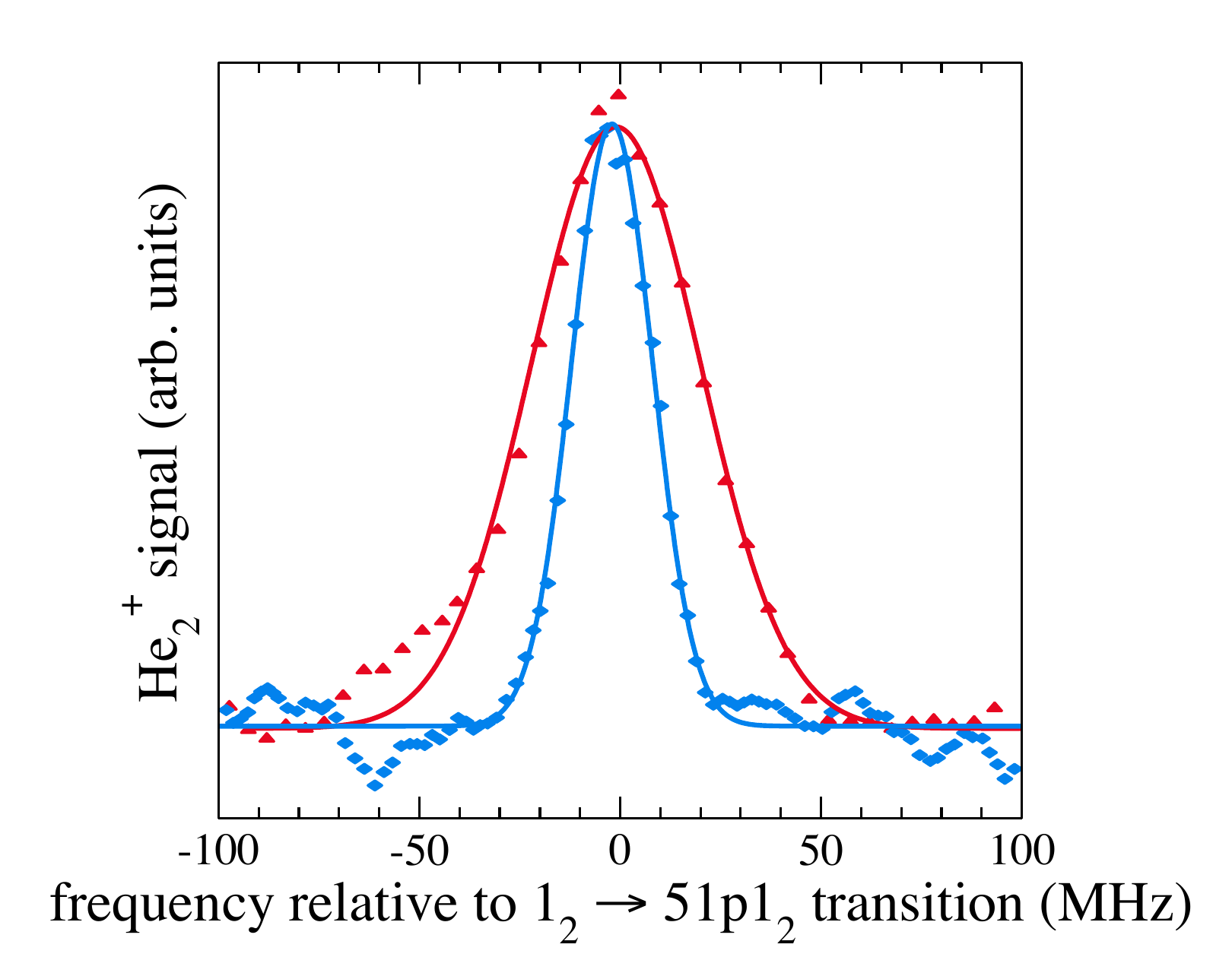}
\caption{Comparison between measurements of the $1_2\rightarrow51\text{p}1_2$ transition using the cw ring dye laser setup of Fig.~\ref{fig:setup}c) with valve temperatures of 77 (solid triangles) and 10\,K (solid diamonds), respectively. The solid red and blue lines represent gaussian fits to the data points with FWHM's of 50, and 25\,MHz, respectively. 
\label{fig:cw_pulsed}}
\end{figure}

As an example, the $1_2\rightarrow 51n\text{p}1_2$ transition, obtained using the cw laser system presented in Fig.~\ref{fig:setup}c), is shown in Fig.~\ref{fig:cw_pulsed}, however, without pulsing currents through the deceleration coils. The red triangles and blue diamonds represent measurements with the valve body kept at a temperature of 77 and 10\,K, respectively. The red and blue curves are Gaussian fits to the data and have a full width at half maximum (FWHM) of 50 and 25\,MHz, respectively. The decrease of the linewidth by a factor of two reflects the reduction in velocity from 1000 to 500\,m/s for the lower temperature. It is expected that decelerating the molecules to velocities around 100\,m/s will result in a further reduction of the observed linewidths, and work in this direction is currently in progress. 

\section{Conclusions}
High-resolution spectroscopy of high Rydberg states of He$_2$ using a  pulsed UV laser with a Fourier-transform-limited bandwidth of 150 MHz was used to determine energy intervals between rotational states of He${_2}^+$, the $N^+=1$ to $N^+=3$ interval with a precision of 18 MHz and the $N^+=11$ to the $N^+=13$ interval with a precision of 150 MHz. Both intervals are smaller than predicted by the most recent \emph{ab initio} calculations \cite{tung2012-1}, which do not include QED corrections to the level energies.

Multistage Zeeman deceleration was used to generate slow beams of translationally cold metastable He$_2$ molecules (He$_2^*$) and contributed to a reduction of the uncertainties in the experimental transition frequencies by reducing Doppler shifts. A complete redistribution among near-degenerate magnetic sublevels of the spin-rotational levels of metastable He$_2$ in regions of near-zero magnetic fields located in the pumping sections of the decelerator was shown to affect the relative populations of the $N^{\prime\prime},J^{\prime\prime}$ spin-rotational states of He$_2^*$, and, by doing so, facilitated the spectral assignments.

Replacing the pulsed UV laser by a cw UV laser enabled us to reduce the line widths of the observed transitions from 150 MHz to 50 and 25 MHz in experiments carried out with He$_2^*$ beams having average velocities of 1000 and 500 m/s, respectively.  The combination of multistage Zeeman deceleration of He$_2^*$ with cw-laser excitation has the potential to improve the precision and accuracy of the present results by more than an order of magnitude, as recently demonstrated using an ultracold sample of cesium atoms in Ref.~\cite{deiglmayr2015}. 

\section*{Acknowledgment}
We thank Hansj\"urg Schmutz and Josef Agner for their expert technical assistance. This work is supported financially by the Swiss National Science Foundation under Project No. 200020-159848 and the NCCR QSIT. P. J. acknowledges ETH Zurich for support through an ETH fellowship.

\end{document}